\documentclass[journal]{IEEEtran}

\usepackage{cite}
\usepackage{amsmath,amsfonts}
\usepackage{algorithmic}
\usepackage{graphicx}
\usepackage{textcomp}
\usepackage{xcolor}
\usepackage{float}
\usepackage{url}
\usepackage{booktabs,threeparttable}
\usepackage{array}
\usepackage{multirow} 
\usepackage{comment}
\usepackage[caption=false,font=normalsize,labelfont=sf,textfont=sf]{subfig}
\usepackage{pifont}
\usepackage{balance}

\def\BibTeX{{\rm B\kern-.05em{\sc i\kern-.025em b}\kern-.08em
    T\kern-.1667em\lower.7ex\hbox{E}\kern-.125emX}}

\newcommand{\ourFramework}{D-MUTRA}

\newcommand{\cmark}{\checkmark}
\newcommand{\xmark}{\texttimes}
  
\begin{document}

\title{\LARGE \ourFramework{}: DLT-based MUTual Remote Attestation for Multi-Agent Systems}

\author{
Adam Zahir,
Vincent Lefebvre,
Mark Angoustures,
Milan Groshev,
and~Carlos J. Bernardos% <-this % stops a space
\thanks{Adam Zahir and Carlos J. Bernardos are with the Department of Telematics, Universidad Carlos~III de Madrid, 28911 Leganes, (e-mail: azahir@pa.uc3m.es; cjbc@it.uc3m.es)}%
\thanks{Vincent Lefebvre and Mark Angoustures are with sarl TAGES SOLIDSHIELD, 06110 Le Cannet, France (e-mail: vincent@solidshield.com; mark@solidshield.com)}%
\thanks{Milan Groshev is with the IE School of Science and Technology, 28046, Madrid, Spain (e-mail: milan.groshev@ie.edu)}%
}

\maketitle

\begin{abstract}
Multi-agent systems (MAS) comprise autonomous software agents that collaborate to perform complex tasks in cyber-physical domains, including multi-robot coordination and the Industrial Internet of Things (IIoT). In such distributed environments, a compromised agent may execute modified software while appearing trustworthy, causing other agents to act on false information and corrupting the mission. Agents must therefore establish and maintain mutual trust throughout operation. Remote attestation (RA) is a well-established technique for this purpose, enabling a remote agent to assess the integrity of a potentially compromised agent. However, conventional RA approaches face significant limitations in MAS: integrity guarantees are restricted to boot or application-load time, designs rely on centralized trusted verifiers or security hardware, and attestation records lack transparency and auditability. To address these limitations, this paper presents \ourFramework{}, a blockchain-based framework that introduces a mutual RA protocol in which agents measure their runtime integrity while verifying that of their peers, acting as both prover and verifier. The framework operates entirely in software and relies on two components: a Security-as-a-Service that instruments agents with lightweight measurement and verification capabilities, and a smart contract that coordinates the attestation protocol in a decentralized and transparent manner. We implement a proof-of-concept on a private Ethereum blockchain using Hyperledger Besu and evaluate it in a swarm robotics scenario built with Robot Operating System (ROS) and the Gazebo simulator. Results show that \ourFramework{} enables agents to continuously attest one another, detects malicious software modifications, and scales to large deployments with negligible overhead on protected applications.
\end{abstract}


\section{Introduction}
\label{sec:intro}

Multi-Agent Systems (MAS) are composed of autonomous software agents that collaborate to solve complex problems in cyber-physical domains, such as multi-robot systems, the Industrial Internet of Things (IIoT), and autonomous network management~\cite{mas-survey}. In such distributed environments, a compromised agent may execute modified software while appearing trustworthy, causing other agents to act on false information and corrupting the mission. Ensuring that agents execute the intended software is therefore a fundamental security requirement in MAS~\cite{mas-security-survey}, commonly addressed through \emph{remote attestation} (RA), a mechanism that enables a remote entity, the \emph{verifier}, to assess the integrity of a potentially compromised device or software agent, the \emph{prover}~\cite{RA-principles-coker}. However, conventional RA provides only a point-in-time guarantee, typically at boot or application loading, leaving runtime modifications undetected~\cite{RA-hw-cflat}. \emph{Runtime integrity verification} (RIV) addresses this limitation by continuously measuring software integrity during execution, extending RA assurances throughout the agent's operational lifetime.

Despite extensive research on RA and RIV~\cite{collective-attestation-SEDA,collective-attestation-SANA,collective-attestation-DARPA,collective-attestation-RADIS,collective-attestation-ScaRR,collective-attestation-kucab,blockchain-RA-javaid,blockchain-RA-BARRET,blockchain-RA-DAN}, existing schemes rely on a trusted verifier to evaluate the integrity of remote devices. However, MAS operate as peers and must continuously verify each other's integrity, as compromises may occur at any point during execution. We define this trust model as \emph{mutual RA}, in which agents act as both prover and verifier and sustain this mutual verification continuously through RIV. This continuous, mutual verification must remain lightweight, as frequent measurements strengthen security but add computational overhead that impacts the performance of the protected applications~\cite{lacoste2023trusted}. Realizing mutual RA further requires tamper-resistant infrastructure that operates without a central authority while ensuring secure, transparent, and immutable interactions. Distributed Ledger Technologies (DLTs) such as \emph{blockchain} naturally satisfy these requirements. Although prior work has explored decentralizing RA through collaborative verification schemes~\cite{collective-attestation-PASTA, collective-attestation-DIAT} and blockchain~\cite{blockchain-RA-PERMANENT,blockchain-RA-LegIoT,blockchain-RA-zRA,blockchain-RA-SCRAPS,blockchain-RA-PONTIS}, these approaches assume a unidirectional trust model and depend on dedicated security hardware, such as a Trusted Platform Module (TPM) or Trusted Execution Environment (TEE) (e.g., Intel SGX or ARM TrustZone). To the best of our knowledge, no existing work supports mutual RA through blockchain technology.

To address this gap, we propose \ourFramework{} (\textbf{D}LT-based \textbf{MUT}ual \textbf{R}emote \textbf{A}ttestation): a blockchain-based framework for mutual RA in MAS. \ourFramework{} allows distributed agents to continuously measure their own integrity at runtime while simultaneously verifying that of their peers. The framework is implemented entirely in software without hardware dependencies and relies on two key components: a Security-as-a-Service (SECaaS) backend that equips agents with lightweight measurement and verification capabilities before deployment, and a blockchain smart contract that coordinates the protocol in a decentralized and auditable manner. The smart contract dynamically assigns the verifier role to the most recently attested agent, creating a chain of trust in which every verifier is the last verified participant.

In summary, the contributions of our research are as follows:
\begin{itemize}
\item We introduce \ourFramework{}, a blockchain-based framework for mutual RA in MAS that enables agents to continuously prove their own integrity and verify that of their peers at runtime, applicable to any software application running on commodity hardware.
\item We implement a working prototype of \ourFramework{} on a private Ethereum blockchain using the open-source Hyperledger Besu platform.
\item We experimentally evaluate \ourFramework{} in a swarm robotics scenario, demonstrating its performance, scalability, and security resilience.
\item We provide a security analysis covering common attack vectors, including software tampering, impersonation, and denial-of-service attacks.
\end{itemize}

The rest of the paper is organized as follows. 
Section~\ref{sec:background} provides the necessary background to understand the concepts discussed in the paper.
Section~\ref{sec:challenges} outlines the key challenges of mutual RA in MAS.
Section~\ref{sec:framework} describes the \ourFramework{} architecture and workflows.
Section~\ref{sec:security-analysis} analyzes its security properties.
Sections~\ref{sec:evaluation} and~\ref{sec:results} present the experimental evaluation and results.
Section~\ref{sec:discussion} discusses the results, limitations, and future work.
Section~\ref{sec:soa} compares related work.
Finally, Section~\ref{sec:conclusions} concludes the paper.
\section{Background}
\label{sec:background}

\subsection{Multi-Agent Systems}

Multi-Agent Systems (MAS) are composed of software programs (agents) that cooperate within a shared environment to execute tasks that would be impractical for a single agent to perform alone. By distributing computation and decision-making, MAS improve scalability, flexibility, and fault tolerance. Representative applications include multi-robot systems, the IIoT, and autonomous network management~\cite{mas-survey}.

The distributed and autonomous characteristics that make MAS effective also introduces security concerns. Compromised agents may disrupt system behavior, manipulate shared information, or influence collective decisions~\cite{mas-security-survey}. Before exchanging data or collaborating, agents must establish mutual trust by verifying two properties: $i)$~\emph{identity}, that a peer is the entity it claims to be, and $ii)$~\emph{software integrity}, that its executable code and runtime state have not been maliciously modified. While identity can be established through standard authentication mechanisms such as public-key cryptography, verifying software integrity requires a security mechanism known as \emph{remote attestation} (RA).

\subsection{Remote Attestation}

RA enables a remote entity to assess over a network whether the software running on a device has been modified or deviates from a trusted configuration~\cite{RA-principles-coker}. A typical RA protocol involves two parties: the \emph{verifier}, which evaluates attestation evidence, and the \emph{prover}, the device under inspection. The prover generates cryptographic measurements of its software through a Root of Trust (RoT) component, which provides a trusted foundation for the measurement process. Attestation may be initiated by the verifier through a challenge-response protocol to obtain fresh measurements from the prover or performed periodically. The prover sends the resulting evidence to the verifier, which compares it against known-good reference measurements to assess device trustworthiness. RA schemes are generally classified into three categories.

\emph{Hardware-based} solutions~\cite{ima, RA-hw-atrium} equip the prover with a dedicated security component, such as a TPM~\cite{RA-hw-tpm} or a TEE (e.g., Intel SGX~\cite{RA-hw-intel-sgx} or ARM TrustZone~\cite{RA-hw-arm-trustzone}), to generate and sign attestation evidence. These components provide tamper-resistant key storage and strong integrity guarantees. However, dependence on vendor-specific hardware limits portability and increases deployment costs.

\emph{Software-based} solutions~\cite{RA-sw-pioneer, RA-sw-swatt} eliminate hardware dependencies by relying entirely on software mechanisms, such as cryptographic hashing of program memory, to verify device integrity. These approaches suit legacy devices and heterogeneous hardware environments but typically require strong adversarial assumptions, including direct communication between prover and verifier, and lack secure storage for cryptographic keys or attestation code.

\emph{Hybrid} approaches~\cite{RA-hybrid-vrased, RA-hybrid-smart} combine software techniques with lightweight hardware components such as Read-Only Memory (ROM) and Memory Protection Units (MPU), providing stronger security guarantees than purely software-based solutions with minimal hardware support. These schemes are commonly studied in resource-constrained IoT devices, which may limit their applicability to general-purpose systems.

Although definitions vary slightly, RA is commonly viewed as a mechanism for verifying software integrity at a specific point in time, typically during boot or application loading. As a result, modifications introduced after attestation, such as runtime memory injections or code-reuse attacks, may remain undetected~\cite{RA-hw-cflat}. \emph{Runtime integrity verification} (RIV) addresses this limitation by continuously monitoring software integrity during execution~\cite{collective-attestation-kucab}, which is particularly important for long-running systems.

\subsection{Blockchain}

Supporting RA and RIV in distributed MAS requires tamper-resistant infrastructure that operates without a central authority and ensures attestation interactions remain secure, transparent, and immutable. \emph{Blockchain} naturally satisfy these requirements. Originally introduced as the technology underlying Bitcoin~\cite{bitcoin}, blockchain is a decentralized ledger replicated across nodes in a peer-to-peer network. The ledger consists of an ordered sequence of timestamped blocks that store transactions and arbitrary data. Each block references the cryptographic hash of its predecessor, creating a tamper-resistant chain extending to the genesis block (block 0). New blocks are added through a consensus mechanism that ensures a consistent view of the ledger across nodes. 

Blockchain networks can be classified along two orthogonal dimensions: accessibility and consensus participation. Network accessibility may be \emph{public}, where ledger data are openly accessible to any entity, or \emph{private}, where access is restricted to authorized participants. Consensus participation may be \emph{permissionless}, allowing any participant to validate transactions without prior authorization, or \emph{permissioned}, restricting this role to approved entities.  Public blockchains are typically permissionless and private blockchains permissioned, though hybrid configurations exist.
In \ourFramework{}, we employ a private, permissioned blockchain in which smart contracts coordinate the attestation protocol in a decentralized manner. The immutable ledger maintains a tamper-resistant history of attestation results, enabling agents to make informed trust decisions before collaborating.

% Blockchain networks are commonly classified as \emph{permissionless} (public), where any node can join and participate in consensus, or \emph{permissioned} (private), where participation and data access are restricted to authorized entities.
\section{Mutual Remote Attestation Challenges in Multi-Agent Systems}
\label{sec:challenges}

Existing RA schemes (reviewed in Sec.~\ref{sec:soa}) establish trust in one direction only, with a single trusted verifier assessing the integrity of a remote device. In MAS, however, agents operate as peers and must verify each other's software integrity before collaborating. This bidirectional trust model, where each participant acts as both prover and verifier, is referred to as \emph{mutual remote attestation} (mutual RA). Moreover, integrity cannot be guaranteed at deployment time alone, as compromises may occur at any point during execution. Continuous RIV is therefore a necessary complement to mutual RA throughout each agent's operational lifetime. We identify the following challenges to realize mutual RA in MAS:

% Despite their complementary roles, the integration of RA and RIV remains largely unexplored.

\begin{itemize}
    \item \textbf{C1~--~Decentralized Verification.} 
    Traditional RA protocols rely on a single trusted verifier. In MAS, designating a central trusted authority is impractical, as it introduces a single point of failure vulnerable to denial-of-service (DoS) attacks and offers no guarantee the verifier itself has not been compromised. The challenge is twofold: selecting which agent acts as verifier for each attestation round, and ensuring that the selected agent is itself trustworthy at that time.
    
    \item \textbf{C2~--~Auditable Trust History.} 
    Conventional RA bases trust decisions on the most recent attestation result provided by a designated verifier. A successful attestation confirms only that a device was in a trustworthy state at that specific point in time, with no record of prior compromises. Given the long-term operation of MAS agents, meaningful trust assessment requires a shared, tamper-resistant record of attestation outcomes accessible to all participants.

    \item \textbf{C3~--~Hardware Assumptions.} 
    MAS agents are typically deployed on commodity hardware lacking dedicated security components such as TPMs, TEEs (e.g., SGX, TrustZone), or ROM. However, existing RA schemes largely assume such hardware and are designed for constrained IoT devices with static configurations, limiting their applicability to general-purpose MAS deployments. These limitations highlight the need for a hardware-independent RA mechanism.

    \item \textbf{C4~--~Workload Performance Impact.} 
    Continuous RIV requires each agent to measure its executable code at regular intervals. The measurement frequency determines the vulnerability window, i.e., the period during which a compromise may remain undetected, and affects application performance~\cite{lacoste2023trusted}. Higher frequencies reduce the window but increase computational overhead, while lower frequencies reduce overhead at the cost of a wider window. Balancing this trade-off between security and performance is a practical design challenge.

    \item \textbf{C5~--~Ledger Storage Scalability.}
    In blockchain-based attestation schemes, each verification is typically recorded as a transaction on the ledger. Because RIV generates measurements continuously across multiple agents, transaction volume increases with measurement frequency and system size. Preserving auditability while controlling ledger growth becomes difficult as the system scales.
\end{itemize}
\section{The \ourFramework{} Framework}
\label{sec:framework}

\ourFramework{} addresses the challenges in Sec.~\ref{sec:challenges} through three design decisions. First, to enable decentralized mutual RA and maintain an auditable trust history (C1, C2), the framework leverages a blockchain network and smart contract. Blockchain provides cryptographic integrity guarantees and an immutable record of attestation interactions, while the smart contract coordinates protocol execution transparently and assigns verifier responsibilities to the most recently verified agent. Second, to eliminate hardware dependencies (C3), attestation is implemented entirely in software, enabling deployment across heterogeneous platforms. Third, to bound the overhead of continuous RIV on protected workloads and the blockchain storage (C4, C5), the framework uses a configurable measurement interval to limit the computational load on each agent and a rolling hash mechanism to aggregate multiple measurements into a single transaction submission, decoupling blockchain growth from measurement frequency. The following subsections describe the \ourFramework{} system architecture (Sec.~\ref{subsec:system-architecture}) and operational workflows (Sec.~\ref{subsec:workflows}).

\subsection{System Architecture}
\label{subsec:system-architecture}

\begin{figure}[t!]
   \centering    
   \includegraphics[width=\columnwidth]{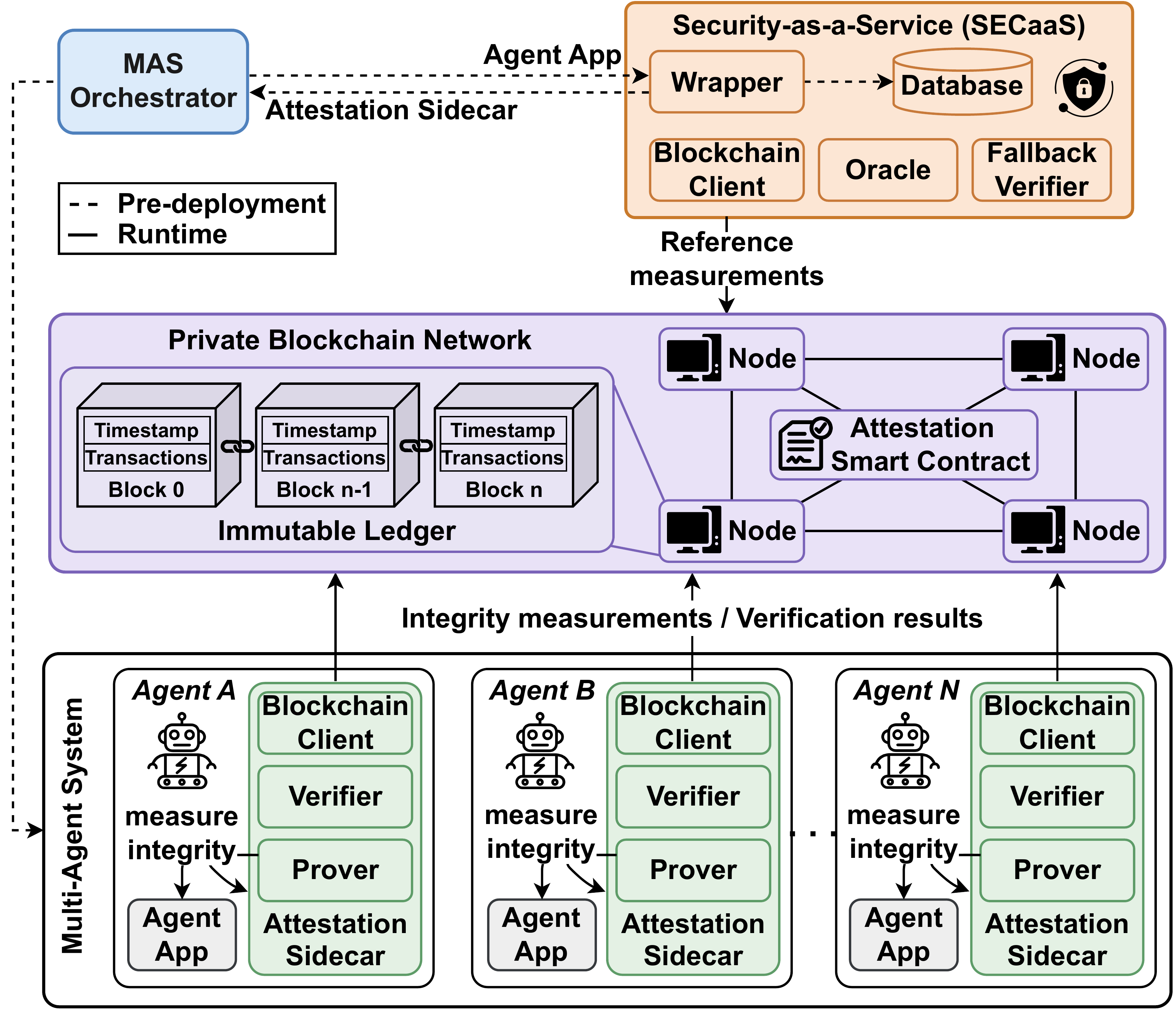}
    \caption{\ourFramework{} architecture supporting mutual remote attestation in MAS.} 
    %\caption{\ourFramework{} architecture supporting mutual remote attestation in MAS. The framework comprises four components: the MAS agents, untrusted software entities undergoing mutual integrity verification; the MAS orchestrator, which manages agent deployment and configuration; SECaaS, which prepares agents for attestation and provides reference measurements for runtime verification; and the blockchain network, which coordinates the protocol and maintains an immutable record of outcomes.} 
      \label{fig:system-architecture}
      \vspace{-3mm}
\end{figure}

Fig.~\ref{fig:system-architecture} presents the \ourFramework{} architecture, comprising four components: the \emph{MAS agents}, untrusted entities undergoing mutual integrity verification; the \emph{MAS orchestrator}, which manages agent deployment and configuration; \emph{Security-as-a-Service} (SECaaS), which prepares agents for attestation and provides reference measurements at runtime; and the \emph{blockchain network}, which coordinates the protocol and maintains a tamper-resistant, auditable record of all outcomes.

\textbf{MAS Agents.}
Each agent (lower part of Fig.~\ref{fig:system-architecture}) consists of its native application and a lightweight \emph{attestation sidecar} process that runs alongside it. The sidecar enables each agent to prove its own integrity and verify that of its peers, supporting mutual RA through three modules: the prover, the verifier, and the blockchain client. 

The \emph{prover} computes integrity measurements and submits them to the smart contract. At startup, it hashes the executable code of both the application and sidecar to establish a trusted baseline, then repeats this measurement periodically to detect runtime modifications. Two configurable parameters govern this process. The \emph{sidecar sleep period} (SSP) specifies the interval between successive measurements, setting the trade-off between the vulnerability window and the computational overhead imposed on the agent. To limit the frequency of attestation transactions submitted to the blockchain, consecutive measurements are aggregated into a rolling hash before submission. The \emph{iteration queue size} (IterQ) defines the number of measurements included in each aggregated submission. 

The \emph{verifier} validates the integrity of other agents. When selected by the smart contract, it retrieves the prover's measurement and the corresponding SECaaS reference, reconstructs the expected rolling hash, compares it with the submitted value, and records the verification result on the blockchain.

The \emph{blockchain client} provides a lightweight interface between the agent and the blockchain. To minimize overhead, particularly on resource-constrained edge devices, agents do not host blockchain nodes or participate in the consensus protocol. Instead, the client signs and submits transactions on behalf of the agent and forwards smart contract events to the prover and verifier. Notably, the sidecar interacts with the rest of the system exclusively through blockchain transactions, keeping it decoupled from the core agent application.

\textbf{Blockchain Network.}
\ourFramework{} targets MAS deployments operating within a single administrative domain, such as industrial multi-robot systems and IIoT infrastructures. In these environments, the attestation protocol must restrict participation to authorized entities while supporting low-latency processing for timely detection of compromised agents. To meet these requirements, we employ a private and permissioned blockchain that provides controlled participation and predictable, low-latency transaction processing. The network (center of Fig.~\ref{fig:system-architecture}) comprises distributed nodes that maintain the shared ledger, order and validate transactions through a consensus protocol, and execute smart contract logic deterministically. These nodes operate independently of agent workloads and may be deployed on dedicated infrastructure within or outside the MAS environment. Every transaction is cryptographically signed, timestamped, and permanently recorded. At the core of the network, the \emph{attestation smart contract} autonomously coordinates the attestation protocol. It receives measurements from provers, selects the most recently attested agent as the verifier, grants it access to the SECaaS reference measurement, and records the verification result on the ledger, thereby maintaining a sequential chain of trust.

%The blockchain network (center of Fig.~\ref{fig:system-architecture}) is the trusted coordination layer of \ourFramework{}. Distributed nodes maintain the shared ledger, order and validate transactions through a consensus protocol, and execute smart contract logic deterministically. These nodes operate independently of agent workloads and can be deployed within or outside the MAS environment. Every transaction is cryptographically signed, timestamped, and permanently recorded. Given the target deployment scenarios (e.g., multi-robot systems, IIoT), we adopt a permissioned blockchain network to maintain high security levels and restrict participation to authorized entities. At the core of the network, the \emph{attestation smart contract} coordinates the attestation protocol. It receives measurements from provers, selects the most recently attested agent as verifier, grants it access to the SECaaS reference measurement, and records the verification outcome on the ledger, establishing a sequential chain of trust.

\textbf{Security-as-a-Service (SECaaS).}
SECaaS (upper-right of Fig.~\ref{fig:system-architecture}) supports \ourFramework{} before deployment and runtime phases. Before deployment, the \emph{wrapper} generates an attestation sidecar, computes a reference measurement by hashing the combined application and sidecar code, and creates a unique blockchain key pair per agent. These artifacts are stored in a secure \emph{database} accessible only to the \emph{oracle}, which supplies reference measurements to the smart contract, and the \emph{fallback verifier}, which performs verification when no peer agent is available. At runtime, the oracle monitors the smart contract for attestation requests, retrieves the corresponding reference measurement, and submits it to the blockchain while restricting access to the verifier selected for that round. If no eligible peer agent is available, the fallback verifier performs the attestation by comparing the agent's measurement with the stored reference and submitting the result to the smart contract.

\textbf{MAS orchestrator.}
The MAS orchestrator (upper-left of Fig.~\ref{fig:system-architecture}) manages agent provisioning and deployment. It submits each agent application to the SECaaS wrapper to obtain an attestation-ready image, deploys each agent together with its sidecar, and configures SSP and IterQ according to the deployment requirements.

\subsection{Operational Workflows}
\label{subsec:workflows}

\ourFramework{} operates as an evolving chain of trust. In each attestation round, an agent proves its integrity to the most recently attested peer, which serves as its verifier. Upon successful attestation, the agent assumes the verifier role for the next round. This process is realized through three workflows: \emph{pre-deployment}, which prepares agents for attestation and registers them with the smart contract; \emph{initial attestation}, which establishes the initial trust anchor for newly deployed agents; and \emph{continuous attestation}, which maintains trust through ongoing integrity verification.

% \ourFramework{} operates through three workflows: \emph{pre-deployment}, which prepares agents for attestation and registers them in the smart contract; \emph{initial attestation}, which establishes the first trust anchor when an agent starts up; and \emph{continuous attestation}, in which agents continuously verify each other's integrity throughout operation.

\textbf{Pre-deployment.}
This phase is executed offline by the MAS orchestrator and only needs to be done once. For each agent, the orchestrator submits the application to SECaaS, which generates an attestation sidecar, computes a reference integrity measurement $h^{\mathrm{ref}}$, and creates a unique blockchain address and private key. The private key enables the agent to sign transactions, while the blockchain address serves as its public identifier. The reference measurement and blockchain address are stored in the SECaaS database, and the address is registered with the smart contract to authorize the agent to submit attestation transactions.

\textbf{Initial Attestation.}
When a newly deployed agent starts up, it immediately performs its initial attestation. This process follows the same procedure as continuous attestation (described below) except that the prover submits a single measurement rather than a hash computed from successive measurements (i.e., a rolling hash), and SECaaS acts as the fallback verifier.

\begin{figure}[tb]
   \centering    
\includegraphics[width=1\columnwidth]{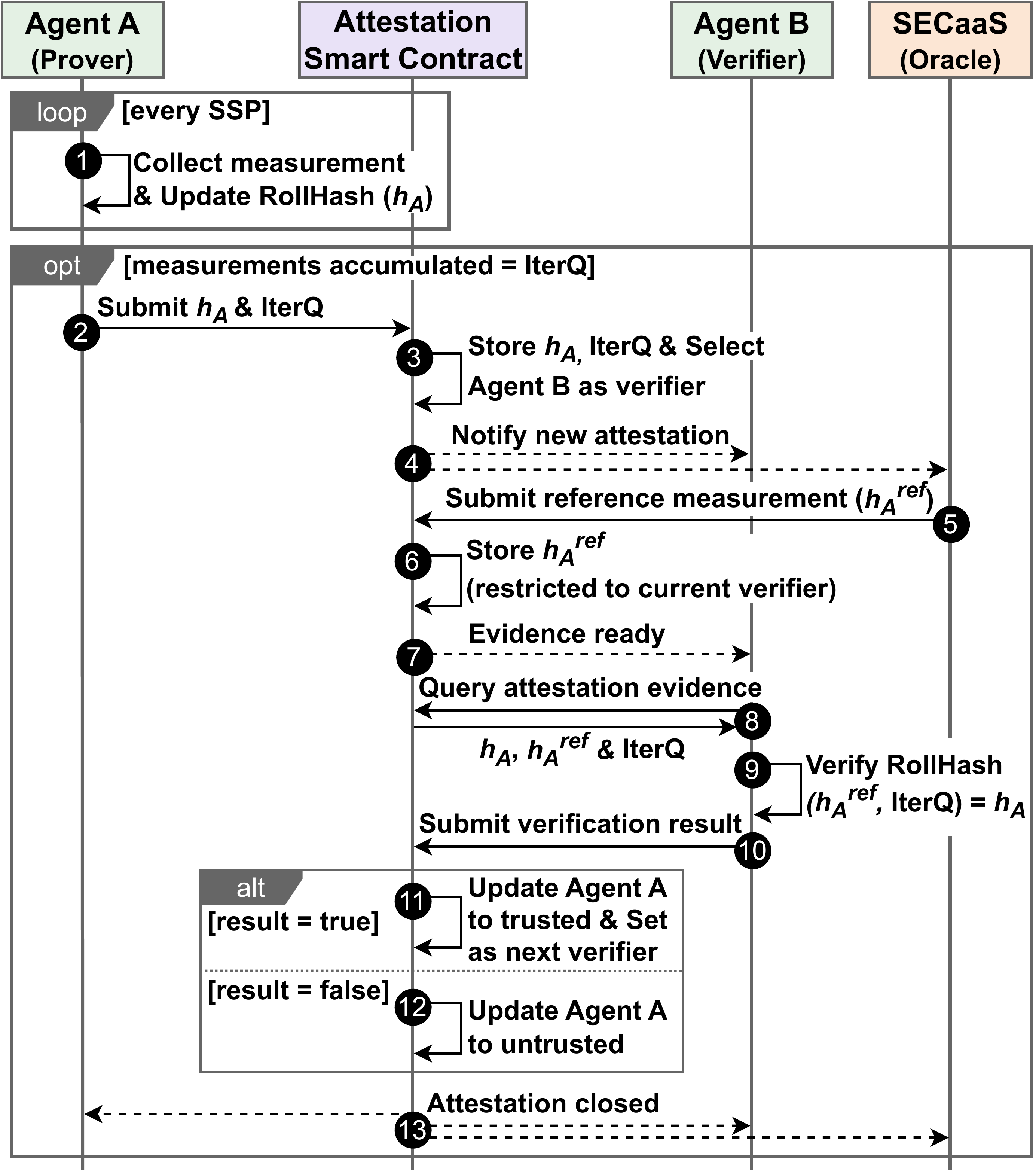}
      \caption{Continuous attestation workflow.}
      \vspace{-5mm}
      \label{fig:runtime-attestation}
\end{figure}

\textbf{Continuous Attestation.}
Once initial attestation succeeds, each agent participates in continuous attestation throughout its operational lifetime. Fig.~\ref{fig:runtime-attestation} illustrates a representative round, with agent~$A$ as the prover and agent~$B$, the most recently verified agent, as the verifier. 
At each SSP interval, agent~$A$ (sidecar process) collects an integrity measurement and updates a rolling hash (step~1). After IterQ measurements, it submits the rolling hash, $h_A$ and IterQ to the smart contract (step~2). The smart contract stores the submitted data, assigns agent~$B$ as verifier, and emits an event announcing a new attestation round (steps~3--4). SECaaS retrieves the corresponding reference measurement $h_A^{\mathrm{ref}}$ from its database and submits it to the smart contract (steps~5--6). It is important to note that access to the reference measurement is restricted to the designated verifier, and the identities of both the prover and verifier remain hidden from all other agents until the attestation round is completed. Agent~$B$ is then notified that the verification evidence is ready (step~7). It retrieves $h_A$, $h_A^{\mathrm{ref}}$, and IterQ (step~8), reconstructs the expected rolling hash, compares it with $h_A$, and submits the verification result to the smart contract (steps~9--10). On success, agent~$A$ is recorded as trusted and set as the verifier for the next round (step~11); on failure, it is recorded as untrusted and excluded from verifier selection until a future attestation succeeds (step~12). Finally, the smart contract broadcasts an attestation closed event (step~13), disclosing the blockchain addresses of the prover and verifier, the attestation outcome, and timestamp. This on-chain record enables agents to query peer trust status and full attestation history before collaboration.

\section{Security Analysis}
\label{sec:security-analysis}

This section presents the adversary model assumed by \ourFramework{} and analyzes its resilience against the considered threat classes.

\subsection{Adversary Model}
\label{subsec:adversary-model}

We consider an adversary with access to the network environment in which agents operate, capable of observing communications and targeting both agents and the attestation process. The common threat model in the MAS security and RA literature identifies three principal threat classes~\cite{mas-security-survey, collective-attestation-DIAT, collective-attestation-PASTA}:
$i)$~\emph{software tampering}, where the adversary modifies an agent's code or injects malicious instructions to alter runtime behavior;
$ii)$~\emph{impersonation}, where an adversary operates under a false identity to influence attestation outcomes; and
$iii)$~\emph{denial-of-service (DoS)}, where the adversary targets the entities responsible for verification---particularly verifier agents and SECaaS---to disrupt the completion of integrity checks.
Physical attacks on the underlying hardware and attacks against the blockchain consensus mechanism are considered out of scope.

\subsection{Security Analysis}
\label{subsec:security-analysis-details}

$i)$~\emph{\textbf{Software Tampering.}}
\ourFramework{} addresses software tampering through the use of RA and RIV. RA establishes a trusted baseline at deployment by verifying agent integrity before admission, while RIV maintains that baseline during execution. At each SSP interval, the prover measures the application and sidecar executables and incorporates the results into a rolling hash submitted to the smart contract. The designated verifier compares the fresh measurement against the SECaaS reference and records the result on-chain. Any mismatch marks the agent as untrusted and excludes it from verifier selection, detecting runtime modifications that RA alone cannot capture. Verifier trustworthiness is reinforced through rotation: the most recently attested agent is selected as verifier each round, minimizing the time elapsed since its last integrity check.

$ii)$~\emph{\textbf{Impersonation.}}
Agent identity rests on two layers. Before deployment, SECaaS provisions each agent with a blockchain key pair, which must be registered in the smart contract before participation. At runtime, each transaction is validated by the blockchain against the sender's signature and by the smart contract against the registered address set. Identity is further coupled with integrity. The reference measurement covers both application and sidecar binaries and is bound to the agent's blockchain address in the SECaaS database. Impersonation therefore requires simultaneously compromising the private key and replicating the exact software state of the target agent. Network-layer attacks such as ARP spoofing or IP spoofing are not applicable, since agents interact exclusively through blockchain transactions.

$iii)$~\emph{\textbf{Denial-of-Service (DoS).}}
All attestation interactions are mediated through the smart contract, eliminating the direct communication channels that conventional RA schemes expose to DoS attacks. Verifier and prover identities are disclosed only when attestation closes, preventing targeted attacks on specific verifiers. Every interaction requires a signed transaction from a registered address, blocking unauthorized participation. When no eligible verifier agent is available, the SECaaS fallback verifier ensures continuity of the protocol.
\section{Experimental Evaluation}
\label{sec:evaluation}

\subsection{Experimental Setup}
\label{subsec:testbed}
To evaluate \ourFramework{}, we designed a swarm robotics scenario based on multi-robot formation control, a representative MAS application widely studied in the literature~\cite{schranz2020swarm}. The scenario models an industrial inspection task where robots follow predefined trajectories while maintaining a geometric formation to maximize area coverage. To support distributed control, robots periodically exchange position information. A compromised robot may inject falsified position updates, leading to incorrect control actions that disrupt the formation and degrade mission performance. \ourFramework{} enables the early detection and isolation of compromised robots through continuous software attestation.

The experimental testbed (Fig.~\ref{fig:testbed}) consists of two virtual machines (VMs) and one bare-metal server. VM-01 runs the SECaaS and the MAS orchestrator, while VM-02 hosts a private blockchain network, each provisioned with 4~vCPUs and 4~GB of RAM. Robotic agents run in simulation on the bare-metal server, equipped with an Intel Core Ultra 9 275HX CPU (24 cores, up to 6.5~GHz), 64~GB of RAM, and an NVIDIA RTX 5080 GPU. The testbed follows a cloud-native design where all components run as Docker containers, including:

\begin{itemize}
\item \textbf{Blockchain network.} 
The private blockchain is built on Hyperledger Besu\footnote{\url{https://besu.hyperledger.org}}, an open-source framework for enterprise-grade blockchains based on the Ethereum ecosystem. The network consists of four validator nodes using the QBFT (Quorum Byzantine Fault Tolerance) consensus, which tolerates up to one-third ($1/3$) malicious validators and provides fast block finality. The consensus is configured with a 2~s block period, corresponding to the average time to validate transactions, include them in a block, and append the block to the blockchain. This configuration is widely adopted in private Ethereum networks with small validator sets, as it balances transaction-confirmation latency and throughput~\cite{besuPerformance}. The smart contract (\texttt{Attestation.sol}), coded using the Solidity programming language, realizes the attestation protocol described in Sec.~\ref{subsec:workflows} and is deployed on this network. Agents and SECaaS interact with the contract through Web3-compatible JSON-RPC endpoints exposed by the validator nodes.
    
\item \textbf{MAS agents.}
Each agent consists of two containers sharing a PID namespace. The robot application runs on Robot Operating System 2 (ROS~2) and controls a simulated TurtleBot3\footnote{\url{https://github.com/ROBOTIS-GIT/turtlebot3_simulations}} mobile robot executing the formation-control task. Its functionality is implemented through a set of ROS~2 nodes (software modules) written in C++, including: an odometry publisher that estimates and broadcasts robot position; a formation controller that combines local and peer odometry data to compute velocity commands for maintaining the desired geometric formation; and mobility drivers that execute the motion commands. The sidecar, implemented in Go, operates independently of the application and includes prover and verifier modules connected to the blockchain through a \texttt{go-ethereum}\footnote{\url{https://geth.ethereum.org}} client. The prover accesses the robot process through the \texttt{/proc}\footnote{\url{https://docs.kernel.org/filesystems/proc.html}} filesystem, extracts the executable code section, computes a SHA-256 measurement over the application and sidecar binaries, and submits it to the smart contract. The verifier listens to contract events and, when selected, compares fresh and reference measurements before recording the result on-chain.

\begin{figure}[t!]
   \centering    
   \includegraphics[width=1\columnwidth]{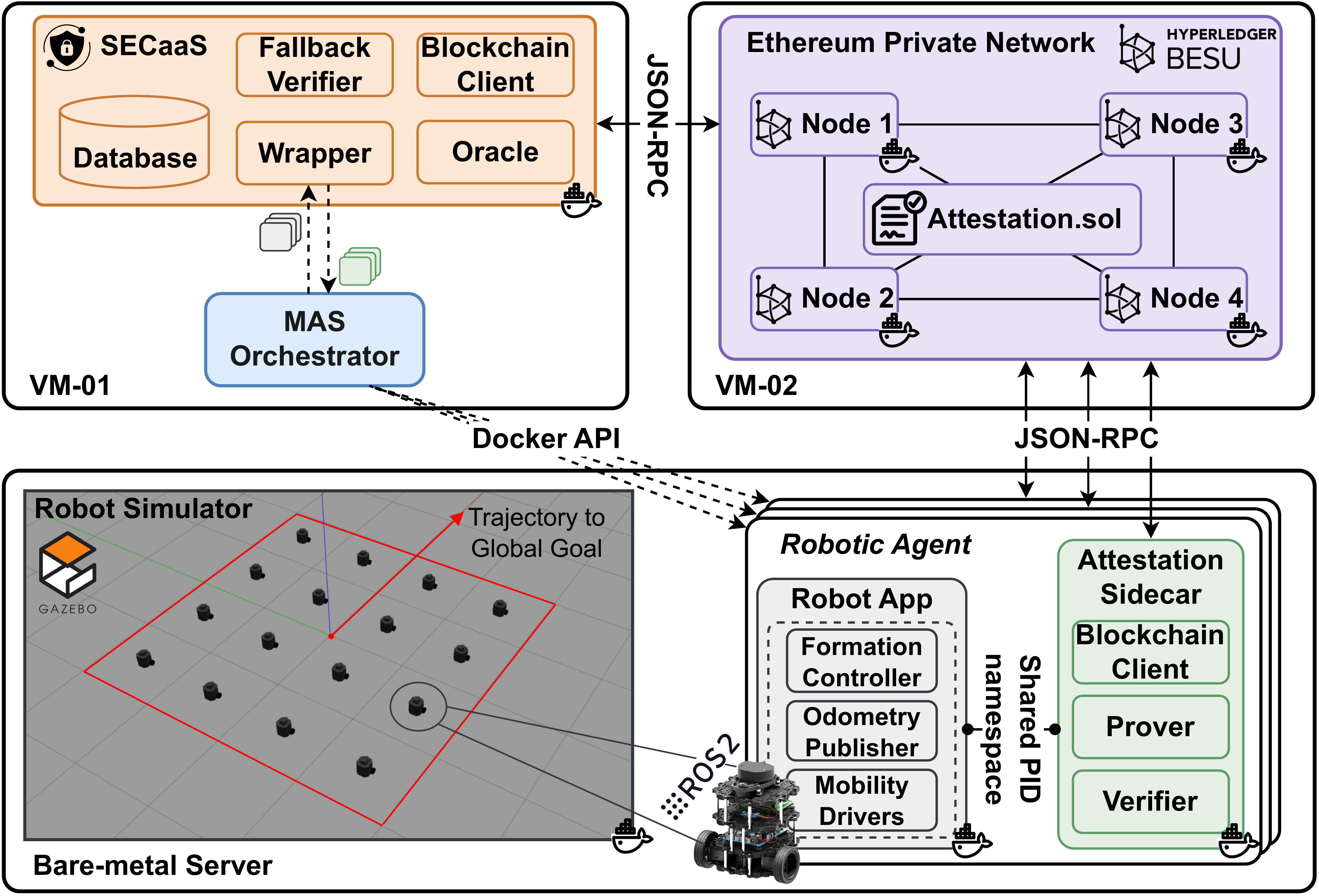}
      \caption{Swarm robotics experimental testbed and scenario.}
      \vspace{-3mm}
      \label{fig:testbed}
\end{figure}

\item \textbf{Robot simulator.} 
The Gazebo\footnote{\url{https://gazebosim.org}} simulator provides a shared backend for all agents, handling physics simulation, sensor data generation, and robot state updates.

\item \textbf{SECaaS.} 
SECaaS hosts five coordinated services that prepare agents for attestation at pre-deployment and support the protocol at runtime. The wrapper exposes a REST API through which the MAS orchestrator submits robot binaries and receives the corresponding sidecar binaries. For each agent, it parses the robot binary in ELF (Executable and Linkable Format) using \texttt{pyelftools}\footnote{\url{https://github.com/eliben/pyelftools}}, generates the sidecar, computes a SHA-256 reference measurement over both binaries, and creates a blockchain key pair. Reference data are stored in a PostgreSQL\footnote{\url{https://www.postgresql.org}} database. At runtime, the oracle and fallback verifier interact with the blockchain via \texttt{web3.py}\footnote{\url{https://web3py.readthedocs.io}} library.

\item \textbf{MAS orchestrator.} 
A Python-based orchestrator coordinates sidecar preparation via SECaaS and manages agent lifecycle operations (creation, modification, deletion) in the container infrastructure.

\end{itemize}

The testbed is designed to emulate a real-world deployment of physical robots and distributed infrastructure for the trusted components of \ourFramework{} (i.e., the MAS orchestrator, SECaaS, and the blockchain network). The robot application stack and attestation sidecar are identical to those deployed on physical hardware, while only the robot dynamics and environment sensing are abstracted by the simulator. Moreover, the co-location of blockchain nodes on a single VM does not introduce artificial performance advantages because the consensus block period (2~s) dominates attestation latency and is several orders of magnitude longer than the millisecond-scale network delays typical between physical nodes in on-premises deployments. Therefore, distributing the blockchain across separate hosts or replacing simulated robots with physical ones would produce equivalent performance results.

\subsection{Evaluation Methodology}
\label{subsec:methodology}

We evaluate \ourFramework{} along two dimensions: performance and scalability under normal operation, and security resilience under a runtime tampering attack.

\textbf{Performance and scalability.} 
We assess the two phases described in Sec.~\ref{subsec:workflows}: \emph{initial attestation} and \emph{continuous attestation}. In the initial phase, each robot performs a single attestation immediately after deployment, with SECaaS acting as the fallback verifier. In the continuous phase, robots periodically submit integrity measurements as provers and, when selected by the smart contract, verify other robots, with SECaaS serving as oracle for reference measurements. Experiments are conducted with swarm sizes of $N \in \{4, \ldots, 64\}$ concurrently attesting robots, using 20 independent runs per configuration. Continuous attestation experiments run for 2~min. Unless otherwise specified, the sidecar sleep period (SSP), i.e., the interval between consecutive measurements, is set to 10~s, and the iteration queue size (IterQ), i.e., the number of consecutive measurements aggregated into a rolling hash before submission, is set to 1. Please note that each attestation cycle comprises three sequential on-chain transactions (prover submission, oracle response, and verifier result). Since each transaction must be confirmed in a separate block before the next can proceed, and the blockchain is configured with a 2~s block period, the minimum theoretical attestation latency is $3 \times 2~\text{s} = 6~\text{s}$. Therefore, the SSP is set to 10~s to provide sufficient timing margin and ensure that each attestation cycle completes before the next measurement is generated.

% The selected SSP therefore provides sufficient margin for to accommodate transaction processing delays.

\begin{itemize}

\item \textbf{Attestation cycle time.} Time to complete a single attestation cycle from the prover's perspective, from the start of SHA-256 digest computation until reception of the on-chain event confirming attestation completion. Timestamps are collected at transaction submission and event reception points in the prover, verifier, and SECaaS.

\item \textbf{Sidecar resource consumption.} CPU (vCPUs) and memory (active working set, MB) usage of the sidecar container. Both metrics are sampled at 1~s intervals via the Docker Engine API\footnote{\url{https://docs.docker.com/reference/api/engine}}, which isolates sidecar computational overhead from other host processes.

\item \textbf{Robot message interval.} Interval between consecutive odometry messages published by the attested robot application, measured by an independent subscriber ROS~2 node. A baseline is established with the sidecar disabled, then measurements are repeated under decreasing SSP values to assess the impact of high-frequency attestation on application timing.

\item \textbf{Blockchain storage growth rate.} Growth rate of the blockchain measured at a single node for different swarm sizes and IterQ values.

\end{itemize}

\textbf{Security resilience.}
We evaluate system resilience using a four-robot swarm in a square formation executing a straight-line trajectory. Three scenarios are considered. In the \emph{baseline}, no attack is present. In the \emph{unmitigated} scenario, NOP instructions are injected into the \texttt{.text} segment of one robot, corrupting its formation controller to produce a constant angular velocity of 2.5~rad/s with zero linear velocity. As a result, the robot spins in place without forward motion. Because robots make control decisions based on odometry from their peers, the failure propagates through the swarm and prevents mission completion. This scenario measures the impact of the attack over a 300~s interval. In the \emph{mitigated} scenario, the same attack is launched with \ourFramework{} enabled. The compromised robot is detected and excluded from the formation control loop, after which the remaining three robots reconfigure into a triangular formation and complete the mission. This scenario evaluates the effectiveness of timely attack detection through attestation. Resilience is quantified using the following metrics:

\begin{itemize}

\item \textbf{Tampering detection time.} Time between runtime tampering and on-chain recording of the corresponding verification failure.

\item \textbf{Swarm formation degradation.} Total formation error accumulated during the compromise window, defined as the interval from attack injection to attack detection (mitigated case) or mission failure (unmitigated case). At each timestep, formation error is computed as the mean Euclidean distance between each robot and its assigned formation position relative to the swarm centroid, capturing how distorted the formation shape is at that instant. Ground-truth robot poses are obtained from the Gazebo physics engine. The cumulative formation error is the sum of these per-timestep errors over the compromise window. 
\end{itemize}

\section{Results}
\label{sec:results}

\subsection{Performance and Scalability Results}

\begin{figure}[tb]
   \centering    
   \includegraphics[width=1\columnwidth]{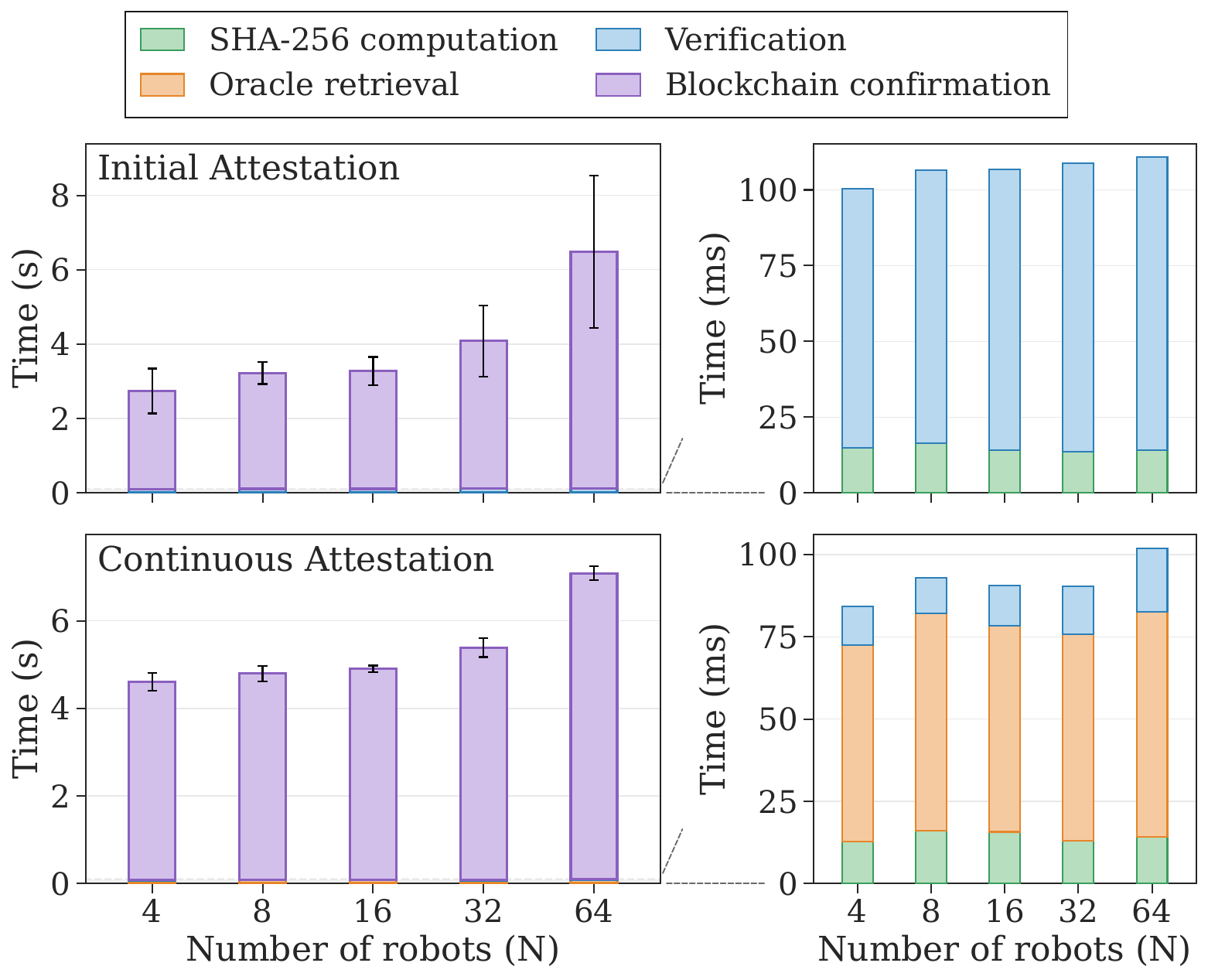}
   \caption{Attestation cycle time as a function of the number of robots ($N$), for initial (top) and continuous (bottom) attestation, decomposed into its main latency components. Bars indicate the mean across robots and runs, while error bars denote the standard deviation.}
   \vspace{-3mm}
    \label{fig:results-attestation-time}
\end{figure}

\textbf{Attestation cycle time.}
Fig.~\ref{fig:results-attestation-time} shows the attestation cycle time as a function of the number of robots ($N$) for both initial (top) and continuous (bottom) attestation. Each bar breaks down into four contributions: SHA-256 digest computation at the prover (green), reference measurement retrieval and on-chain submission by the SECaaS oracle (orange), digest verification at the verifier (blue), and blockchain confirmation time (purple), which is the combined waiting time for all transactions in the cycle to be validated and recorded on the blockchain. The inset reports the first three contributions at millisecond resolution. 

For initial attestation, the mean cycle time increases from 2.7~s at $N=4$ to 6.5~s at $N=64$, a 2.4$\times$ increase despite a 16$\times$ increase in swarm size. The standard deviation also increases with $N$, reaching $\pm$2.0~s at $N=64$. This is because SECaaS serves as the sole verifier for all robots and processes requests sequentially, introducing queueing delays as $N$ grows. Blockchain confirmation dominates latency at 96--98\% across all $N$. This result is expected because each initial attestation requires two sequential transactions (the prover's measurement and the SECaaS verification result), each awaiting at least one block period (2~s) for confirmation. As shown in the inset, SHA-256 computation remains stable at 13--17~ms, while the SECaaS contribution increases slightly from 85~ms to 97~ms.

For continuous attestation, the mean cycle time increases from 4.6~s at $N=4$ to 7.1~s at $N=64$. The higher baseline compared with initial attestation is due to an additional transaction in which the SECaaS oracle submits the reference measurement, bringing the total to three sequential confirmations per cycle. Despite this, the variance remains low, with a standard deviation below 0.22~s for all $N$. This is because verification is distributed among robots, eliminating the bottleneck associated with a single verifier. The inset further shows that the non-blockchain contributions remain stable: SHA-256 within 13--17~ms, oracle processing within 60--68~ms, and peer verification within 12--19~ms.

\begin{figure}[tb]
   \centering    
   \includegraphics[width=1\columnwidth]{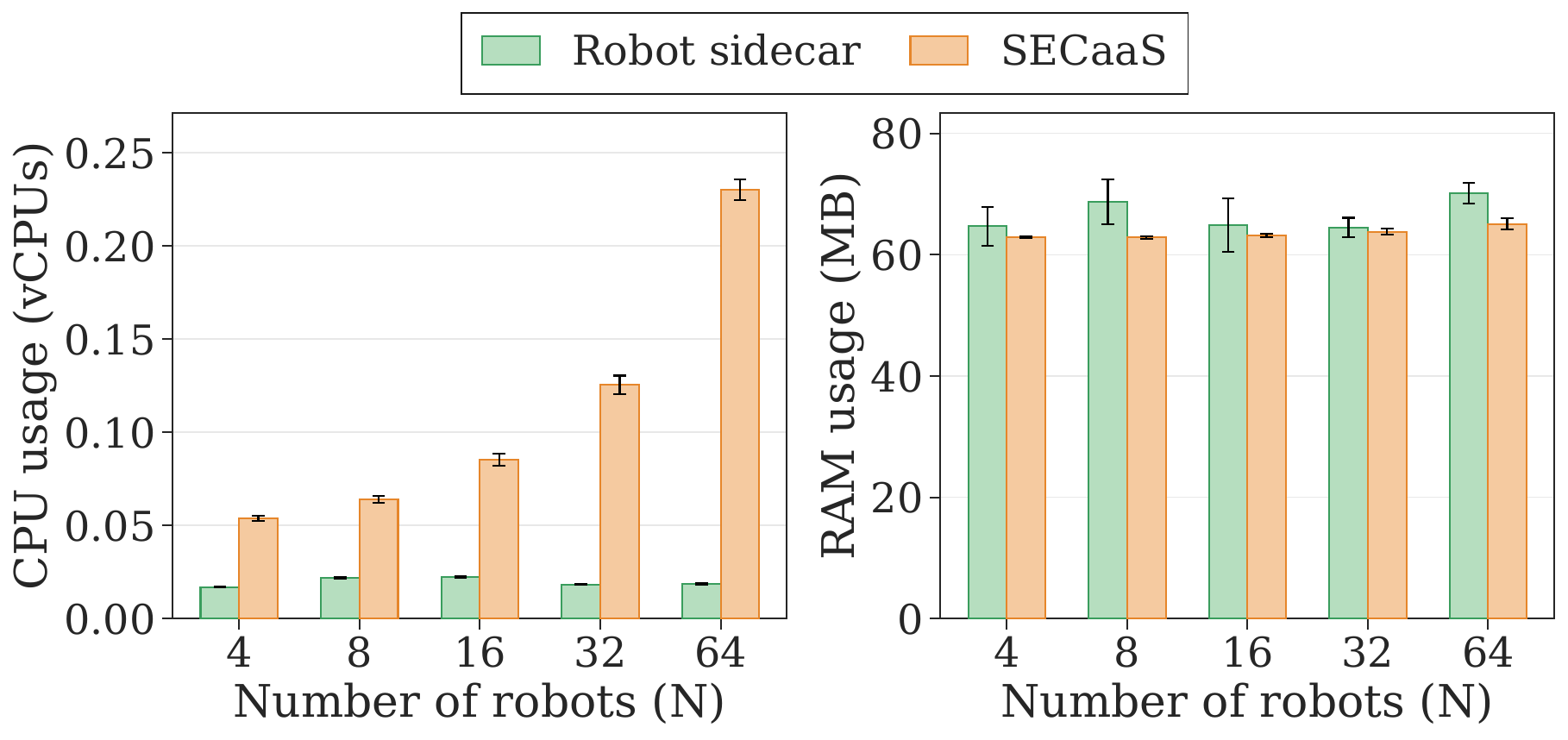}
   \caption{CPU and RAM usage of the sidecar (per robot) and SECaaS during continuous attestation for different swarm sizes ($N$). CPU usage is reported in vCPUs (1 vCPU~=~1 logical core).}
   \vspace{-3mm}
   \label{fig:results-resource-usage}
\end{figure}

\textbf{Resource consumption.}
Fig.~\ref{fig:results-resource-usage} shows the CPU and memory usage of the sidecar (per robot) and SECaaS during continuous attestation. Sidecar CPU remains stable between 0.017 and 0.022~vCPU across swarm sizes ranging from $N=4$ to $N=64$, while memory stays within 65--70~MB. These results indicate negligible scaling overhead and confirm that the sidecar imposes a lightweight, per-robot cost that is effectively independent of $N$. SECaaS CPU usage, on the other hand, increases moderately from 0.054~vCPU at $N=4$ to 0.230~vCPU at $N=64$, reflecting its role as the oracle that serves an increasing number of robots requiring reference measurements. Nevertheless, it remains below 1 vCPU (i.e., one logical core) even at the largest scale. Despite their different roles, both components show similar memory footprints, since memory usage is dominated by the overhead of each runtime environment (Go for the sidecar, Python for SECaaS) rather than by attestation activity itself.

\begin{figure}[tb]
   \centering    
   \includegraphics[width=1\columnwidth]{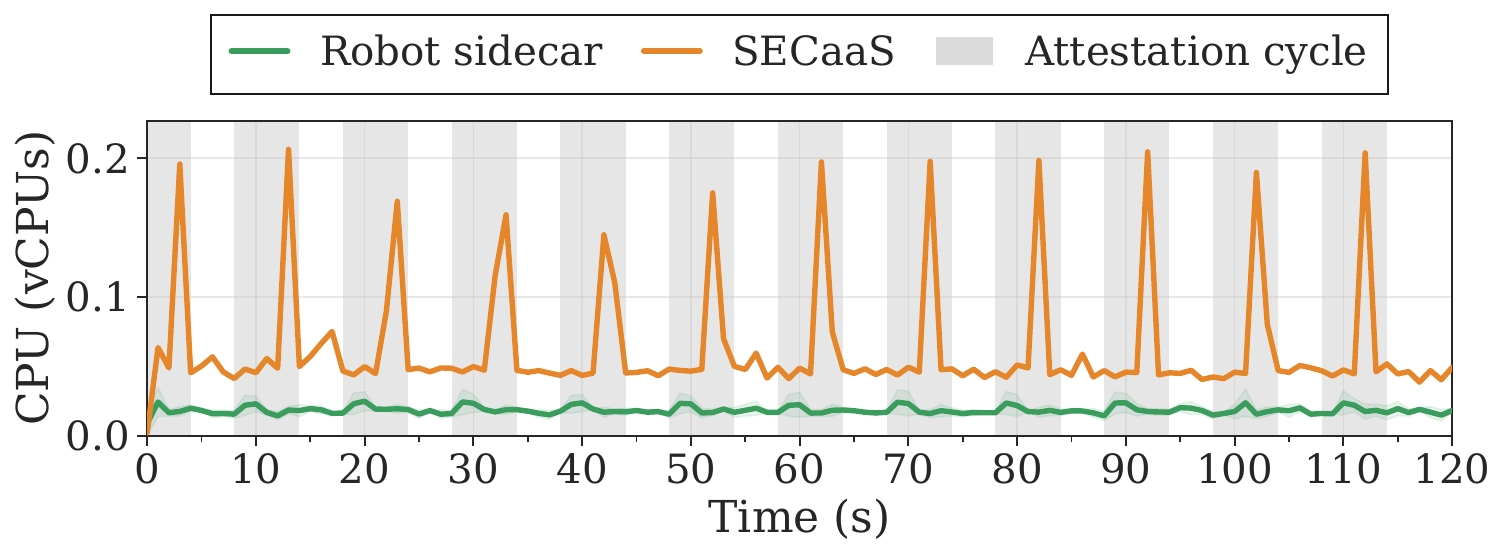}
    \caption{CPU usage of the sidecar (per robot) and SECaaS over time.}
    \vspace{-3mm}
    \label{fig:results-resource-usage-time-series}
\end{figure}

Fig.~\ref{fig:results-resource-usage-time-series} further illustrates the temporal behavior of CPU utilization during a representative run. Both components exhibit periodic bursts of activity every 10~seconds, corresponding to the configured SSP. Notably, The SECaaS CPU peaks are delayed by a few seconds relative to the sidecar, consistent with the sequential execution flow: the oracle processes each request only after the prover's transaction is confirmed on-chain. Between attestation cycles, both components return to low steady-state utilization, confirming that resource consumption is event-driven.

\begin{figure}[t!]
   \centering    
\includegraphics[width=0.95\columnwidth]{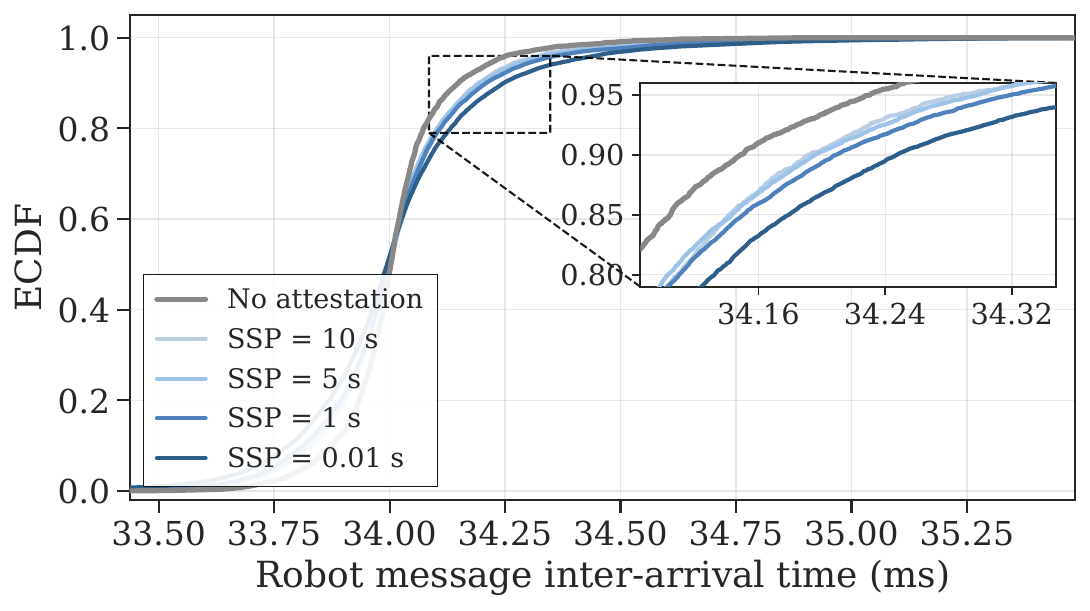}
      \caption{ECDF of message inter-arrival times for the robot application under high-frequency sidecar sleep period (SSP) values.}
      \vspace{-3mm}
    \label{fig:results-robot-performance}
\end{figure}

\textbf{Robot message interval.}
Fig.~\ref{fig:results-robot-performance} shows the empirical cumulative distribution of message inter-arrival times for the attested robot application under different SSP configurations. The median remains constant at 34.0~ms across all settings. Even under the most frequent measurement interval (SSP~=~10~ms), the 99th percentile increases by only 0.34~ms compared to the baseline without attestation. These results indicate that the attestation sidecar introduces no measurable performance overhead on the robot application.

\begin{figure}[t!]
   \centering    
   \includegraphics[width=0.95\columnwidth]{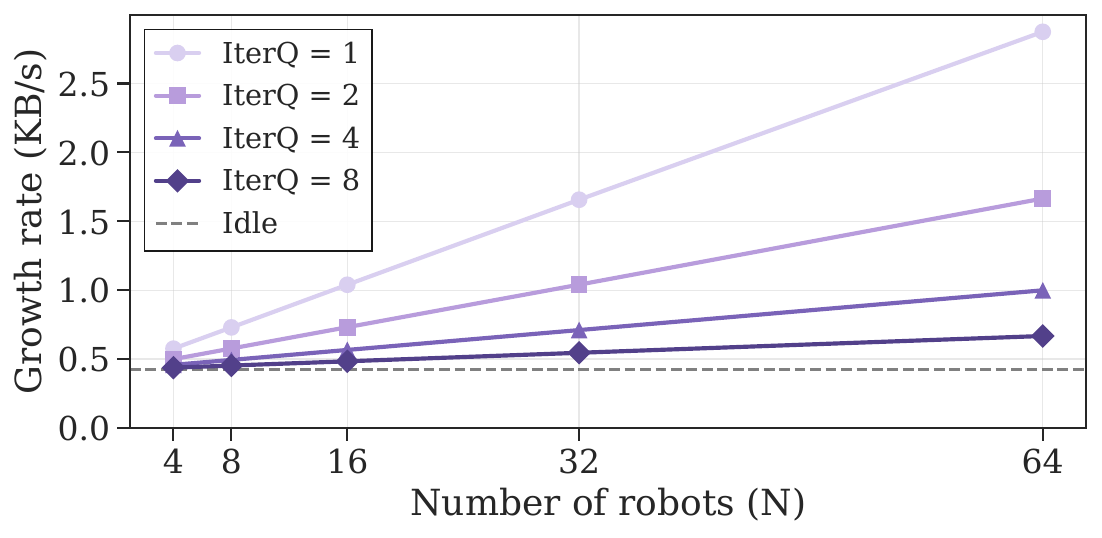}
   \caption{Growth rate of blockchain during continuous attestation for different number of robots ($N$) and iteration queue sizes (IterQ). The dashed line indicates the growth rate without attestation activity.}
    \vspace{-3mm}
    \label{fig:results-blockchain-growth-rate}
\end{figure}

\textbf{Blockchain storage growth.}
Fig.~\ref{fig:results-blockchain-growth-rate} shows the storage growth rate of the blockchain as a function of $N$ for different IterQ values. With IterQ~=~1, each integrity measurement is submitted as a separate transaction, producing the highest growth rate. As $N$ increases from 4 to 64, the growth rate rises from 0.58~KB/s to 2.88~KB/s, a 5$\times$ increase. Increasing IterQ reduces blockchain growth by aggregating consecutive measurements into a single rolling hash before submission. At $N=64$, IterQ~=~4 reduces the growth rate to 1.00~KB/s, a 2.9$\times$ reduction relative to IterQ~=~1. With IterQ~=~8, the growth rate remains nearly constant across all evaluated $N$, and reaches 0.67~KB/s at $N = 64$, remaining close to the idle baseline of 0.42~KB/s measured without attestation activity.

\subsection{Security Resilience Results}

\begin{figure}[t!]
   \centering      \includegraphics[width=1.0\columnwidth]{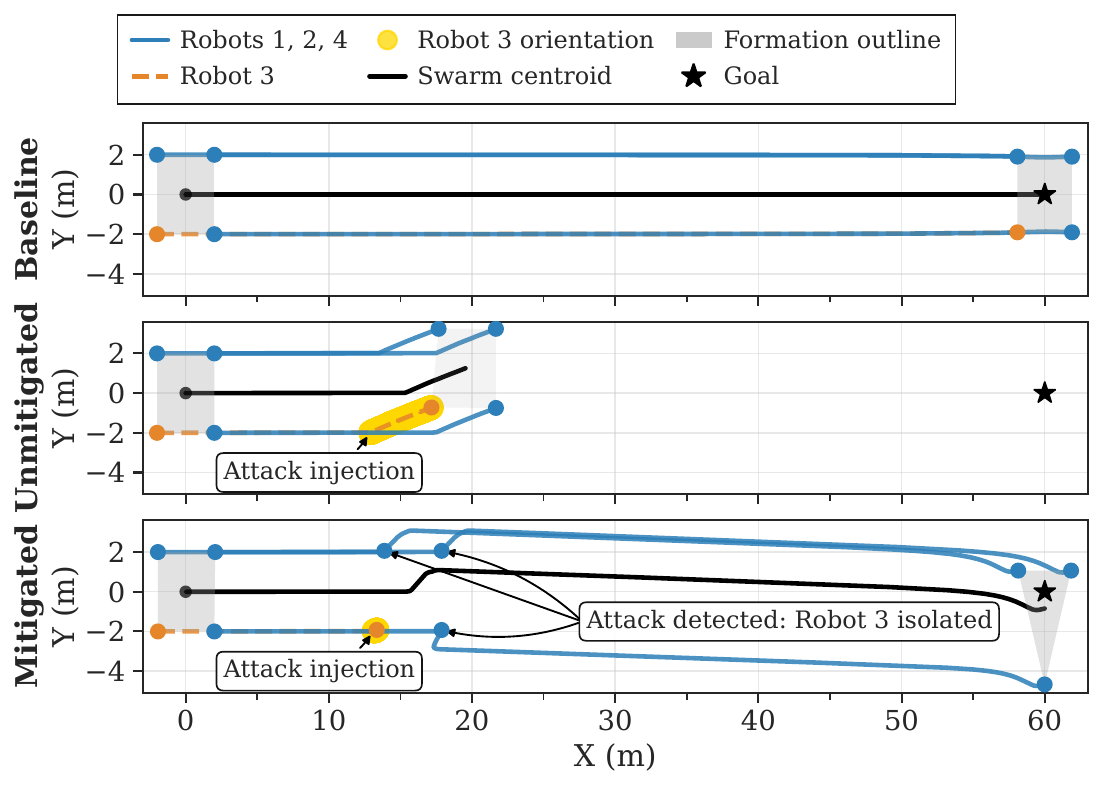}
  \caption{Robot trajectories during the formation-control task under normal operation (top), with an undetected attack (middle), and with the attack detected by \ourFramework{} and the compromised robot isolated (bottom).}
  \label{fig:swarm-trajectory}
\end{figure}

\textbf{Swarm trajectories.}
Fig.~\ref{fig:swarm-trajectory} illustrates the trajectories executed by the four robots under the three experimental conditions. In the baseline case (top), all robots maintain the square formation and successfully reach the goal. Under the unmitigated attack (middle), Robot~3 begins spinning immediately after attack injection, as indicated by its circular trajectory and orientation markers. Because the remaining robots rely on its falsified odometry, the formation becomes unstable and the swarm reaches only 32.6\% of the target distance. With \ourFramework{} enabled (bottom), the compromise is detected, Robot~3 is excluded, and the three healthy robots reconfigure into a triangular formation. The swarm then completes the mission despite the attack.

\begin{figure}[t!]
  \centering
\includegraphics[width=1.0\columnwidth]{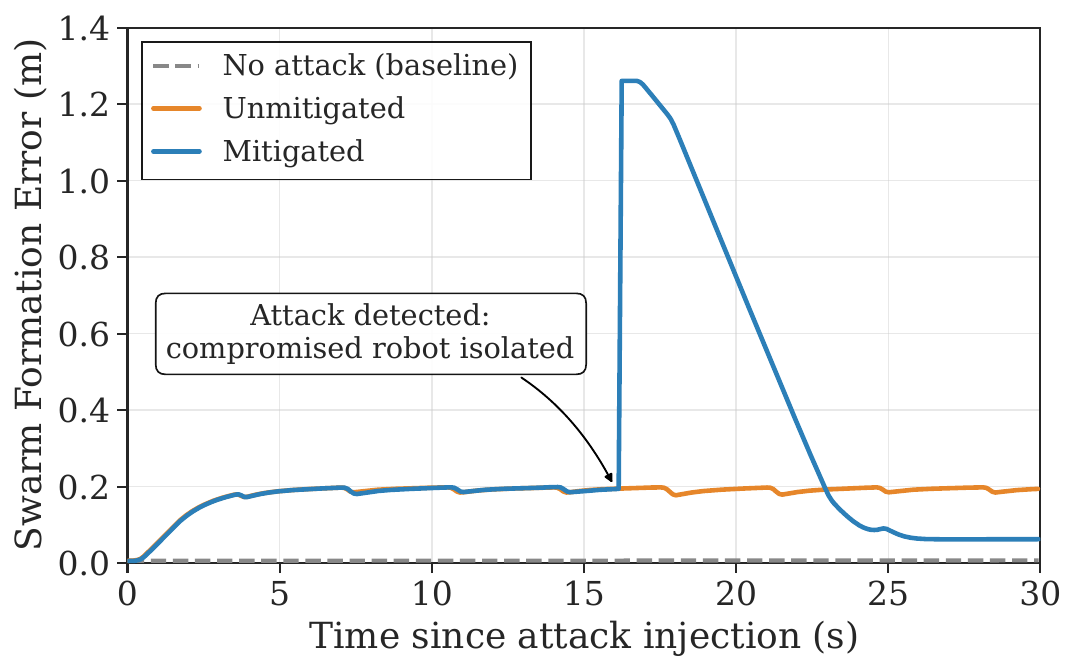}
  \caption{Swarm formation error over time from attack injection, without and with \ourFramework{} enabled. The dashed line indicates the error level under normal operation.}
  \vspace{-3mm}
  \label{fig:results-swarm-error}
\end{figure}

\textbf{Formation error over time.}
Fig.~\ref{fig:results-swarm-error} shows the instantaneous swarm formation error, i.e., the mean deviation of each active robot from its target position relative to the swarm centroid, over time following attack injection. Under both attack conditions, the error rises to 175--200~mm, approximately $25\times$ the baseline noise level of $\sim$7~mm (dashed line), as the manipulated velocity commands from robot~3 progressively distort the swarm geometry. At 16~s, the attack on robot~3 is detected, and the healthy robots reconfigure from a four-robot square to a three-robot triangle. This causes a transient error peak of 1.26~m because the error is evaluated against the new target positions immediately after robot~3 is excluded, before the remaining robots reach their updated locations. The error then decreases as the controllers drive the robots toward the triangle geometry, stabilizing at approximately 60~mm. This residual remains above baseline because the triangle formation has wider spacing between robots. Without mitigation, the error remains near 0.2~m for the rest of the experiment.

 \begin{figure}[t!]
   \centering    
   \includegraphics[width=1.0\columnwidth]{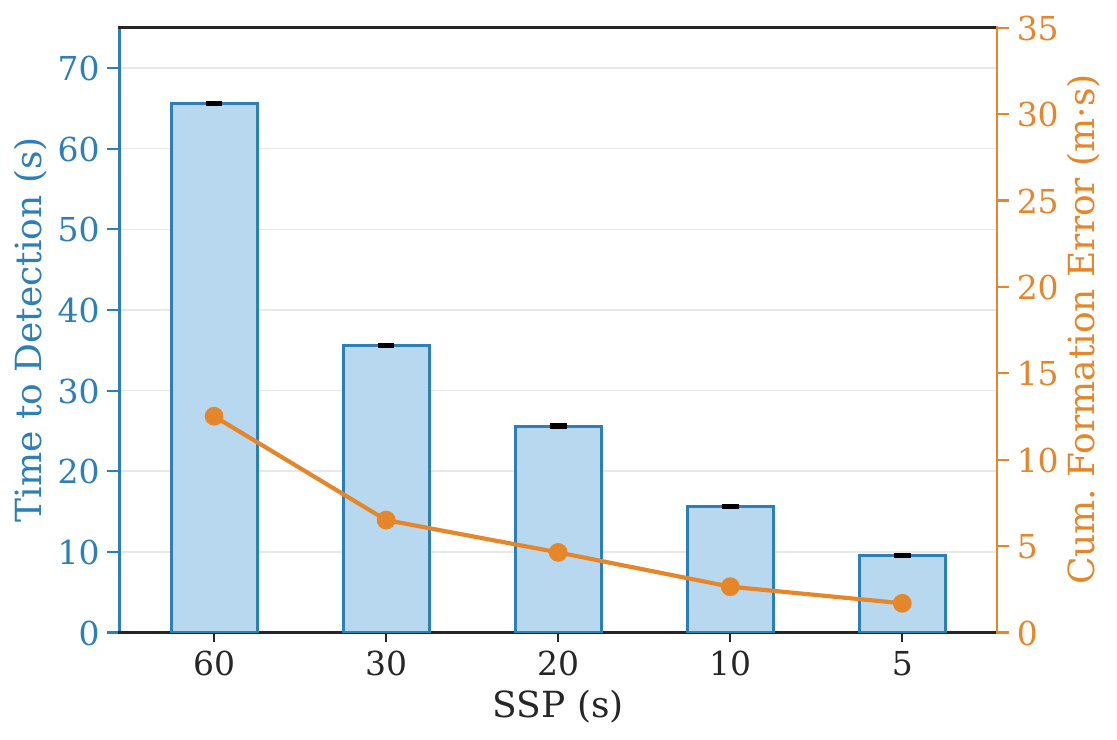}
   \caption{Effect of SSP on tampering detection time (left axis) and cumulative formation error during the compromise window (right axis).}
    \vspace{-3mm}
    \label{fig:results-detection-time}
\end{figure}
%Fig.~\ref{fig:results-swarm-error} shows the swarm formation error over time following attack injection. We can observe that formation error increases identically under both attack conditions to approximately 175--200~mm, about $25\times$ the baseline noise level of $\sim 7$~mm (dashed line). At 16~s, the attack is detected and the healthy robots reconfigure, producing a temporary error peak of 1.26~m. The error then stabilizes at approximately$\sim 60$~mm. This value remains above the baseline because the three-robot formation has a different geometry with larger spacing than the original four-robot formation. In contrast, without mitigation, the formation error remains near 0.2~m for the remainder of the experiment.

\textbf{Tampering detection time and cumulative formation error.}
Fig.~\ref{fig:results-detection-time} shows the effect of SSP on tampering detection time and cumulative formation error during the compromise window. Detection time decreases monotonically with SSP, approximating SSP plus a fixed offset of about 5.6~s. This offset corresponds to the time the blockchain takes to validate the sequence of three transactions involved in a continuous attestation round and to record its outcome, as shown earlier in Fig.~\ref{fig:results-attestation-time}. At SSP~=~60~s, the mean detection time is 65.6~s, while at SSP~=~5~s it reduces to 9.6~s. Cumulative formation damage decreases proportionally, from 12.51~m$\cdot$s to 1.69~m$\cdot$s over the same range, confirming that shorter attestation intervals directly limit mission degradation.
\section{Discussion}
\label{sec:discussion}
The goal of \ourFramework{} is to provide an attestation framework that enables distributed software agents to continuously prove their integrity and verify that of their peers throughout their operational lifetime. It employs a software-based attestation protocol coordinated by a blockchain smart contract, eliminating the need for dedicated security hardware or centralized verifiers. By combining RA with continuous RIV, the framework ensures integrity both at deployment and during runtime. Experimental results demonstrate the feasibility of \ourFramework{}, showing that agents can successfully attest one another, detect malicious modifications, and scale to large deployments with negligible overhead on protected applications.

The evaluation also reveals the costs of this design. Blockchain integration is the primary source of latency, as the protocol relies on multiple transactions that must be propagated, processed, and confirmed before attestation completes. This overhead reflects the cost of implementing attestation on a distributed ledger, where all interactions are recorded as auditable, tamper-resistant transactions. It is worth noting that the average time between two attestations in existing RA schemes is at least 5--10 minutes in the worst cases~\cite{blockchain-RA-zRA}. Consequently, the latency introduced by blockchain communication remains practical for many deployment scenarios.

From a security perspective, attestation latency defines the vulnerability window, i.e., the interval during which a compromise may remain undetected. In \ourFramework{}, this window is determined by the sum of the SSP and the blockchain delay. Its acceptable duration depends on the target application and can be adapted via the SSP. For instance, in the swarm robotics use case evaluated in our experiments, short windows are preferable since compromised agents can rapidly propagate falsified data that degrades mission performance. In contrast, less time-sensitive applications such as IIoT monitoring can tolerate longer detection intervals since the effects of manipulated data typically accumulate gradually.

IterQ introduces an additional trade-off between detection responsiveness and blockchain storage overhead. Aggregating IterQ consecutive measurements into a single rolling hash before submission reduces the number of transactions recorded on-chain. However, each aggregation step delays the publication of attestation evidence by one SSP interval, extending the vulnerability window. For long-term deployments where participating blockchain nodes have limited storage, ledger growth can be further mitigated using established techniques such as state pruning, which removes historical transaction data while preserving the integrity and auditability of the ledger state~\cite{blockchain-pruning}.

% Although \ourFramework{} offers flexibility through configurable security and performance parameters, several limitations remain.

\subsection{Limitations and Future Work}
RIV introduces a Time-of-Check to Time-of-Use (TOCTTOU) gap. During the interval between consecutive measurements, an attacker can modify the executable, run malicious code, and restore the original binary. The rolling hash mechanism partially mitigates this by requiring the binary to be restored before each individual measurement. Nevertheless, more sophisticated attacks remain possible. In a page reconstruction attack, for example, malicious code executes outside the measured memory region while an unmodified replica is maintained in DRAM. A detection gap also exists at agent startup since the blockchain-based attestation process introduces a delay (3--8~s in our experiments) before the initial integrity verdict is established. During this interval, a compromised workload may perform malicious actions before tampering detection. Finally, SECaaS introduces a residual trust dependency on the oracle. If the oracle is unavailable, attestation cannot proceed. If it is compromised, an attacker could manipulate reference measurements and undermine the entire attestation process.

Future work will focus on addressing these limitations in three directions. First, we will distribute the SECaaS backend across multiple independent nodes to eliminate the single point of failure and reduce trust assumptions. Second, we will explore anchoring reference measurements directly on-chain at deployment time, thereby removing reliance on an off-chain oracle. Third, we will investigate synchronized and ephemeral measurement techniques to reduce the TOCTTOU window and strengthen runtime integrity guarantees at the prover level.
\section{Related work}
\label{sec:soa}

\subsection{Collective Remote Attestation}
Collective remote attestation extends single-device verification to networks of devices, enabling scalable integrity verification across multiple peers. Existing schemes either rely on a centralized trusted verifier or distribute verification responsibilities among participating devices.

\textbf{Centralized verifier model.}
SEDA~\cite{collective-attestation-SEDA} organizes provers in a tree topology where attestation evidence propagates upward and is validated by a root verifier. 
SANA~\cite{collective-attestation-SANA} improves on this using Optimistic Aggregate Signatures (OAS) to compress evidence from many provers into a compact proof without trusting intermediate relay nodes.
DARPA~\cite{collective-attestation-DARPA} extends the threat model to physical attacks, detecting compromised devices by monitoring periodic heartbeat exchanges and flagging devices that go offline for a detectable period.
RADIS~\cite{collective-attestation-RADIS} applies this model to distributed IoT services rather than individual devices. Each service attests the control-flow of the service it invokes at runtime, and a single verifier validates the resulting hash chain against a database of legitimate paths computed offline.
Despite their differences, all FOUR schemes rely on a single trusted verifier, creating a single point of failure vulnerable to DoS attacks.

\begin{table*}[t]
\centering
\caption{Comparison of remote attestation schemes, including collective and blockchain-based approaches, against \ourFramework{}.}
\label{tab:related-works}
\resizebox{\textwidth}{!}{%
\begin{tabular}{l l l l l c l}
\toprule
\textbf{Scheme} &
\textbf{Platform} &
\begin{tabular}[c]{@{}l@{}}\textbf{Integrity}\\\textbf{Measurement}\end{tabular} &
\textbf{Verifier} &
\begin{tabular}[c]{@{}l@{}}\textbf{Blockchain}\\\textbf{Role}\textsuperscript{\dag}\end{tabular} &
\begin{tabular}[c]{@{}c@{}}\textbf{Bidirectional}\\\textbf{Trust}\end{tabular} &
\begin{tabular}[c]{@{}l@{}}\textbf{Trust}\\\textbf{Anchor}\textsuperscript{\ddag}\end{tabular} \\
\midrule
\multicolumn{7}{c}{\textit{Collective Remote Attestation}} \\
\midrule
SEDA~\cite{collective-attestation-SEDA}
    & Embedded & Firmware, load-time   & Designated authority  & ---              & \xmark & ROM + MPU \\
SANA~\cite{collective-attestation-SANA}
    & Embedded & Firmware, load-time   & Designated authority  & ---              & \xmark & ROM + MPU \\
DARPA~\cite{collective-attestation-DARPA}
    & Embedded & Firmware, load-time   & Designated authority  & ---              & \xmark & ROM + MPU \\
PASTA~\cite{collective-attestation-PASTA}
    & Embedded & Firmware, load-time   & Peer devices & ---              & \xmark & ROM + MPU \\
DIAT~\cite{collective-attestation-DIAT}
    & Embedded & Control-flow, runtime & Peer devices & ---              & \xmark & TrustZone \\
RADIS~\cite{collective-attestation-RADIS}
    & Embedded & Control-flow, runtime & Designated authority  & ---              & \xmark & None (TrustZone opt.) \\
ScaRR~\cite{collective-attestation-ScaRR}
    & General-purpose & Control-flow, runtime & Designated authority & ---         & \xmark & Kernel (SW) \\
Kucab~\textit{et al.}~\cite{collective-attestation-kucab}
    & General-purpose & Control-flow, runtime & Designated authority & --- & \xmark & SGX + CET \\
\midrule
\multicolumn{7}{c}{\textit{Blockchain-based Remote Attestation}} \\
\midrule
Javaid~\textit{et al.}~\cite{blockchain-RA-javaid}
    & Embedded & Firmware, load-time   & Designated authority  & Record only      & \xmark & PUF \\
BARRETT~\cite{blockchain-RA-BARRET}
    & Embedded & Firmware, load-time   & Designated authority  & Record only      & \xmark & ROM \\
DAN~\cite{blockchain-RA-DAN}
    & Embedded & Firmware, load-time   & Designated authority  & Record only      & \xmark & TPM \\
PERMANENT~\cite{blockchain-RA-PERMANENT}
    & Embedded & Firmware, load-time   & Device itself         & Record only      & \xmark & ROM + RTC \\
LegIoT~\cite{blockchain-RA-LegIoT}
    & Embedded & Firmware, load-time   & Peer devices & Record only      & \xmark & ROM + MPU / TrustZone / TPM \\
zRA~\cite{blockchain-RA-zRA}
    & Embedded & Firmware, load-time   & Any participant     & Verify \& record & \xmark & ROM + MPU \\
SCRAPS~\cite{blockchain-RA-SCRAPS}
    & Embedded & Firmware, load-time   & Proxy smart contract        & Verify \& record & \xmark & None (TrustZone opt.) \\
PONTIS~\cite{blockchain-RA-PONTIS}
    & General-purpose & Enclave image, load-time & Any participant & Verify \& record & \xmark & SGX / TrustZone \\
\midrule
\ourFramework{}
    & General-purpose & Process memory, runtime & Last-attested peer & Coordinate \& record & \cmark & Kernel (SW) \\
\bottomrule
\end{tabular}%
}
\smallskip
\begin{minipage}{\textwidth}
\footnotesize
\textsuperscript{\dag}\textbf{Blockchain Role:}
\textit{Record only} --- the blockchain stores attestation results; verification is performed off-chain.
\textit{Verify \& record} --- the blockchain verifies or distributes attestation evidence.
\textit{Coordinate \& record} --- the blockchain elects verifiers, distributes evidence, and records outcomes.\\
\textsuperscript{\ddag}\textbf{Trust Anchor:} the component protecting attestation code and cryptographic keys on the prover side.
\end{minipage}
\vspace{-3mm}
\end{table*}

\textbf{Distributed verifier model.}
To remove the dependency on a central verifier, several schemes distribute verification responsibility across participating devices.
PASTA~\cite{collective-attestation-PASTA} enables provers to jointly produce attestation tokens via a Schnorr-based multisignature scheme, embedding integrity proofs for all participants into a compact token that any device can verify independently.
DIAT~\cite{collective-attestation-DIAT} targets autonomous systems, where each device verifies the control-flow integrity of a collaborating peer before accepting its data at runtime.
Although these schemes remove the centralized verifier, they require pre-established cryptographic relationships and fixed network topologies with direct connectivity between provers and verifiers. Such assumptions are difficult to satisfy in dynamic environments where devices may join or leave without prior coordination.

\textbf{Attestation of general-purpose systems.}
Most collective attestation schemes target embedded devices with static firmware and predictable memory layouts, while few address general-purpose systems running complex software stacks.
ScaRR~\cite{collective-attestation-ScaRR} provides runtime control-flow attestation for cloud systems via LLVM instrumentation, partitioning execution into sub-paths and verifying them incrementally. Using the Linux kernel as a software trust anchor, it achieves throughput of up to 2 million control-flow events per second. However, compile-time instrumentation introduces runtime overhead, and the design relies on a single verifier. 
Kucab~et al.~\cite{collective-attestation-kucab} combine load-time attestation via Intel SGX with runtime control-flow integrity enforced by Intel CET, avoiding instrumentation with minimal overhead. However, the approach depends on vendor-specific hardware (SGX and CET) and does not support decentralized or mutual verification.

Overall, existing collective attestation schemes adopt a unidirectional trust model in which only the verifier evaluates the prover. Moreover, while some schemes extend attestation to runtime, they rely either on dedicated security hardware or compile-time instrumentation, limiting their applicability across heterogeneous hardware platforms.

\subsection{Blockchain-based Remote Attestation}

Blockchain technology has been explored in RA as a secure infrastructure for recording verification outcomes and enabling decentralized interactions between provers and verifiers. Existing schemes differ in how actively the blockchain participates in the attestation process.

\textbf{Blockchain as record only.}
Several schemes use the blockchain solely as an immutable ledger to store attestation results, while verification is performed off-chain by a designated trusted party.
Javaid~\textit{et al.}~\cite{blockchain-RA-javaid} anchor device identities to Physical Unclonable Functions (PUFs) and store reference measurements on-chain, while a designated verifier challenges provers directly and evaluates responses locally. The blockchain serves only as a data repository, the scheme requires specialized PUF hardware, and provers remain vulnerable to DoS attacks because any entity can issue challenges directly without authorization.
BARRETT~\cite{blockchain-RA-BARRET} mitigates this by requiring attestation requests to be submitted as Ethereum transactions, where each request incurs a gas fee that makes flooding attacks economically prohibitive. Nevertheless, evidence is still evaluated off-chain by a trusted third party.
DAN~\cite{blockchain-RA-DAN} distributes attestation records across organizations using a permissioned blockchain built on Hyperledger Fabric and a consortium of verifiers. While this improves auditability, DAN still relies on TPM hardware on every device and assumes the trustworthiness of the verifier consortium.
PERMANENT~\cite{blockchain-RA-PERMANENT} adopts a self-attestation model in which each device measures its integrity at randomized intervals triggered by a hardware timer and publishes signed results to a permissioned blockchain. Other devices assess trustworthiness by querying on-chain history and computing a weighted trust score. However, integrity is self-declared rather than independently verified, and every device must operate as a full blockchain node, imposing significant computational overhead.
LegIoT~\cite{blockchain-RA-LegIoT} maintains a graph of trust relationships among devices, allowing existing attestation results to be reused across the network and reducing redundant verifications. Attestation is carried out by peer devices, and the blockchain records the resulting trust associations rather than performing verification itself.

\textbf{Blockchain as active participant.}
zRA~\cite{blockchain-RA-zRA}, SCRAPS~\cite{blockchain-RA-SCRAPS}, and PONTIS~\cite{blockchain-RA-PONTIS} use the blockchain to coordinate or verify attestation evidence.
zRA introduces a non-interactive attestation protocol based on zkSNARK proofs, enabling any party to independently verify a device's integrity from its on-chain proof without prior knowledge or a designated verifier. However, integrity measurement remains limited to static firmware, and proof generation introduces non-trivial overhead on the constrained hardware that zRA targets.
SCRAPS delegates verification to a smart contract that evaluates attestation evidence submitted by publishers on behalf of subscribers. This eliminates the need for a trusted verifier and scales to large IoT networks, but retains a unidirectional trust model in which only publishers are attested, while subscribers are never verified.
PONTIS targets general-purpose servers with heterogeneous TEE hardware, using a blockchain-based identity registry to verify TEE instances across different hardware platforms. However, attestation is limited to the enclave image at instantiation time, and every participant requires vendor-specific TEE hardware.

Table~\ref{tab:related-works} summarizes and compares the characteristics of existing RA schemes with those of \ourFramework{}. The comparison shows that \ourFramework{} is the only solution that supports mutual runtime RA on general-purpose platforms without hardware dependencies. It achieves this through a blockchain-coordinated, software-based attestation protocol in which agents continuously measure their runtime integrity while mutually verifying that of their peers.

\section{Conclusions}
\label{sec:conclusions}
This paper presents \ourFramework{}, a novel blockchain-based framework for mutual remote attestation in multi-agent systems. \ourFramework{} enables distributed agents (software applications) to continuously measure their integrity at runtime while verifying that of their peers, acting as both prover and verifier in the attestation protocol. It operates entirely in software, requiring no modifications to protected applications and no security hardware. The framework integrates two components: a Security-as-a-Service backend that prepares agents before deployment with lightweight measurement and verification capabilities, and a blockchain-based smart contract that coordinates the attestation protocol in a decentralized, auditable manner. By designating the most recently attested agent as the next verifier, the contract maintains a dynamic root of trust and reduces the period during which a compromised agent could assume that role. To the best of our knowledge, \ourFramework{} is the first framework to support mutual runtime remote attestation using blockchain technology.

We implemented a proof-of-concept on a private Ethereum blockchain and evaluated in a swarm robotics scenario. Results show that agents can continuously attest one another, detect malicious modifications, and scale efficiently to large multi-agent deployments while imposing negligible overhead on protected applications. These findings confirm that mutual remote attestation provides a practical foundation for establishing and maintaining trust in distributed environments.

\section*{Acknowledgment}
This work was supported in part by the European Union's Horizon Europe research and innovation programme through the NATWORK project under Grant Agreement No. 101139285, which funded the design, development, evolution, and security analysis of \ourFramework{} carried out by Solidshield. The testbed implementation and experimental validation were further supported by the 6GINSPIRE project (PID2022-137329OB-C42), funded by MCIN/AEI/10.13039/501100011033, and by the EURO-3C project, funded by the European Union's Horizon Europe research and innovation programme under Grant Agreement No. 101297599. Views and opinions expressed are however those of the author(s) only and do not necessarily reflect those of all EURO-3C consortium parties nor those of the European Union (granting authority). Neither of them can be held responsible for them.

\ifCLASSOPTIONcaptionsoff
  \newpage
\fi

\vspace{-1em}

\bibliographystyle{IEEEtran}
\bibliography{bibliography}

\vspace{-2.5em}

\begin{IEEEbiographynophoto}{Adam Zahir}
received his M.Sc. in 2024, and is a Ph.D. student at Universidad Carlos III de Madrid (UC3M).
\end{IEEEbiographynophoto}\vspace{-3em}

\begin{IEEEbiographynophoto}{Vincent Lefebvre}
received his M.Sc. in 1988, and is the CEO of sarl TAGES SOLIDSHIELD.
\end{IEEEbiographynophoto}\vspace{-3em}

\begin{IEEEbiographynophoto}{Mark Angoustures}
received his Ph.D. in 2018, and is the CTO at sarl TAGES SOLIDSHIELD.
\end{IEEEbiographynophoto}\vspace{-3em}

\begin{IEEEbiographynophoto}{Milan Groshev}
received his M.Sc. in 2016 and Ph.D. in 2022, and is a postdoctoral researcher at IE University.
\end{IEEEbiographynophoto}\vspace{-3em}

\begin{IEEEbiographynophoto}{Carlos J. Bernardos}
received his M.Sc. in 2003 and Ph.D. in 2006, and is a Professor at UC3M.
\end{IEEEbiographynophoto}

\vfill

\end{document}